\documentclass[aps,pre,showpacs,floats,twocolumn,floats,superscriptaddress,floatfix]{revtex4}

\usepackage{graphicx}
\usepackage{bm}
\usepackage{amsfonts}
\usepackage{color}

\usepackage{amsmath}    
\usepackage{epsfig}
\usepackage{subfigure}  
\usepackage{hyperref}   
\usepackage{bm}
\usepackage{amssymb}

\newcommand{\ocaaddress}{Lab. Lagrange, UCA, OCA, CNRS, CS, 34229, 06304, Nice Cedex 4, France}
\newcommand{\romeaddress}{Department of Physics and INFN, University of 
Rome ``Tor Vergata'',
Via della Ricerca Scientifica 1, 00133, Rome, Italy.}  

\newcommand{\ictsaddress}{International Centre for
  Theoretical Sciences, Tata Institute of Fundamental Research,
  Bangalore 560089, India}

\begin{document}

\title{Intermittency in fractal Fourier hydrodynamics: Lessons from the 
Burgers equation \footnote{Postprint version of the article published on PHYSICAL REVIEW E 93, 033109 (2016)}}

\author{Michele Buzzicotti} 
\email{buzzicotti.m@gmail.com}
\affiliation{\romeaddress}
\author{Luca Biferale} 
\email{biferale@roma2.infn.it}
\affiliation{\romeaddress}
\author{Uriel Frisch} 
\email{uriel@oca.eu}
\affiliation{\ocaaddress}
\author{Samriddhi Sankar Ray}
\email{samriddhisankarray@gmail.com}
\affiliation{\ictsaddress}
\begin{abstract}
We present theoretical and numerical results for the one-dimensional 
stochastically forced Burgers equation decimated on a fractal Fourier set of dimension $D$.

We  investigate the robustness of the energy transfer mechanism and of the small-scale statistical fluctuations 
 by changing $D$. We find that a very small percentage of
mode-reduction  ($D \lesssim 1$) is enough to destroy most of the
characteristics of the original non-decimated equation. In particular,
we observe a suppression of intermittent fluctuations for $D <1$ and a
quasi-singular transition from the fully intermittent ($D=1$) to the
non-intermittent case for $D \lesssim 1$. Our results indicate that the
existence of strong localized structures  (shocks) in the one-dimensional Burgers
equation is the result of highly entangled correlations amongst {\it all} Fourier
modes. 

\end{abstract}
\maketitle

\section{Introduction}
\label{sec:Intro}
An outstanding challenge of the past few decades has been to develop a
rigorous understanding of the  energy transfer from large to small
scales in fully developed, three-dimensional, incompressible turbulent
flows~\cite{frischbook}. Numerical simulations and experiments
 show that  multi-point velocity correlation functions are intermittent, i.e., they
develop a power-law behavior with non-dimensional (anomalous) scaling
exponents \cite{frischbook,K41,grisharmp}. 
The question of the origins of intermittency in turbulence
and its relation {\it inter alia} to small structures  is also of central importance in the areas of
non-equilibrium statistical physics, fluid dynamics, astrophysics, and geophysics \cite{ruelle,vassilicos,astro1,astro2,astro3,astro4,geo1,geo2,math1,math2,math3,math4}.

In this paper, we investigate the small-scale properties of  the 
stochastically-forced one-dimensional Burgers equation which, is a
paradigmatic example of a highly intermittent system with  statistics dominated by coherent structures in physical space (shocks)
\cite{burgulence, becreview}. For this, we perform a series of {\it numerical experiments}
by studying the evolution of the original partial differential equation restricted on a fractal set of Fourier modes \cite{frisch2012}. The idea is to 
reduce the number of degrees of freedom with minimal breaking of the original symmetries of the system. The goal is to understand what are the key
 ingredients in the dynamics necessary to reproduce the main statistical properties of the original non-decimated 
Burgers equation and therefore to  understand the robustness and origins of its shock-like structures.

Over the past few decades various models of intermittency have been developed based on the idea
of energy cascade in physical space~\cite{frischbook}.  These
contrast with other attempts, based on the spectral space, 
involving statistical closures and renormalization group methods. Despite
their success for certain problems, none of these attempts have been  able to
quantitatively connect  anomalous scaling with the structure of the original equation, and hence with its intermittent behavior. 
As a result, our understanding of anomalous scaling is still based 
on phenomenological real-space descriptions and real-space methodologies.

Intermittency is intimately connected with ideas of energy transfer from large
to small scales. Working in Fourier space should, hopefully, open new and  possible ways to understand this cascade.  In a recent work~\cite{frisch2012},
the idea of {\it fractal decimation} was
introduced, with the aim of studying the evolution of the Navier-Stokes equations  on an effective dimension $D$ out of  an integer
$d$-dimensional embedding manifold.  This is done by introducing a quenched mode-reduction 
 in Fourier space such that in a sphere of radius $k$ the number of modes that are
involved in the dynamics scale as $k^D$ (where $D < d$ is the effective fractal dimension 
of the system) for large
$k$~\cite{frisch2012,rayreview}. This approach allows us to decimate the number
of triad interactions in Fourier space as a function of the wavenumbers involved  as well as to
consider the problem in non-integer, fractal dimensions $D$. In Ref.\cite{luca2015} the
first results for a set of simulations of the decimated, three-dimensional (3d)
Navier-Stokes equation have been reported with the intriguing conclusion
that fractal Fourier decimation leads, rather quickly (i.e., for a very small
reduction of the Fourier modes $D\lesssim 3$), to vanishing intermittency. 

In the present paper we investigate the same problem for the one-dimensional
Burgers equation. The main advantage with respect to the previous attempt on
the 3D Navier-Stokes is that here, due to the simpler structure of the
problem, numerical simulations can reach much higher resolutions and therefore
assess, in a fully quantitative way, the problem of scaling and corrections to
it. 

The rest of the paper is organised as follows. In Sec.~\ref{sec:Description} we
introduce the decimated Burgers equation and give details about our numerical
simulations. We then present results from our numerical simulations as well as
provide theoretical and phenomenological arguments to substantiate them. In
particular, in Sec.~\ref{sec:SecondOrder} we examine the effect of
decimation on second-order correlation functions both in Fourier space, via the
energy spectra (Sec.~\ref{subsec:ScalingFourier}), and in physical space
(Sec.~\ref{subsec:ScalingReal}) through the second-order structure function and
flatness. We then, in Sec.~\ref{sec:StrucFunc}, investigate in detail -- by using  numerical
simulations and theory -- the suppression of intermittency by examining the
higher-order structure functions. Finally in
Sec.~\ref{sec:Conclusions} we make concluding remarks and provide a plausible
theoretical framework in which to understand the spectral scaling seen in our
simulations.

\begin{figure*}
\begin{center}
\includegraphics[width=0.497\linewidth]{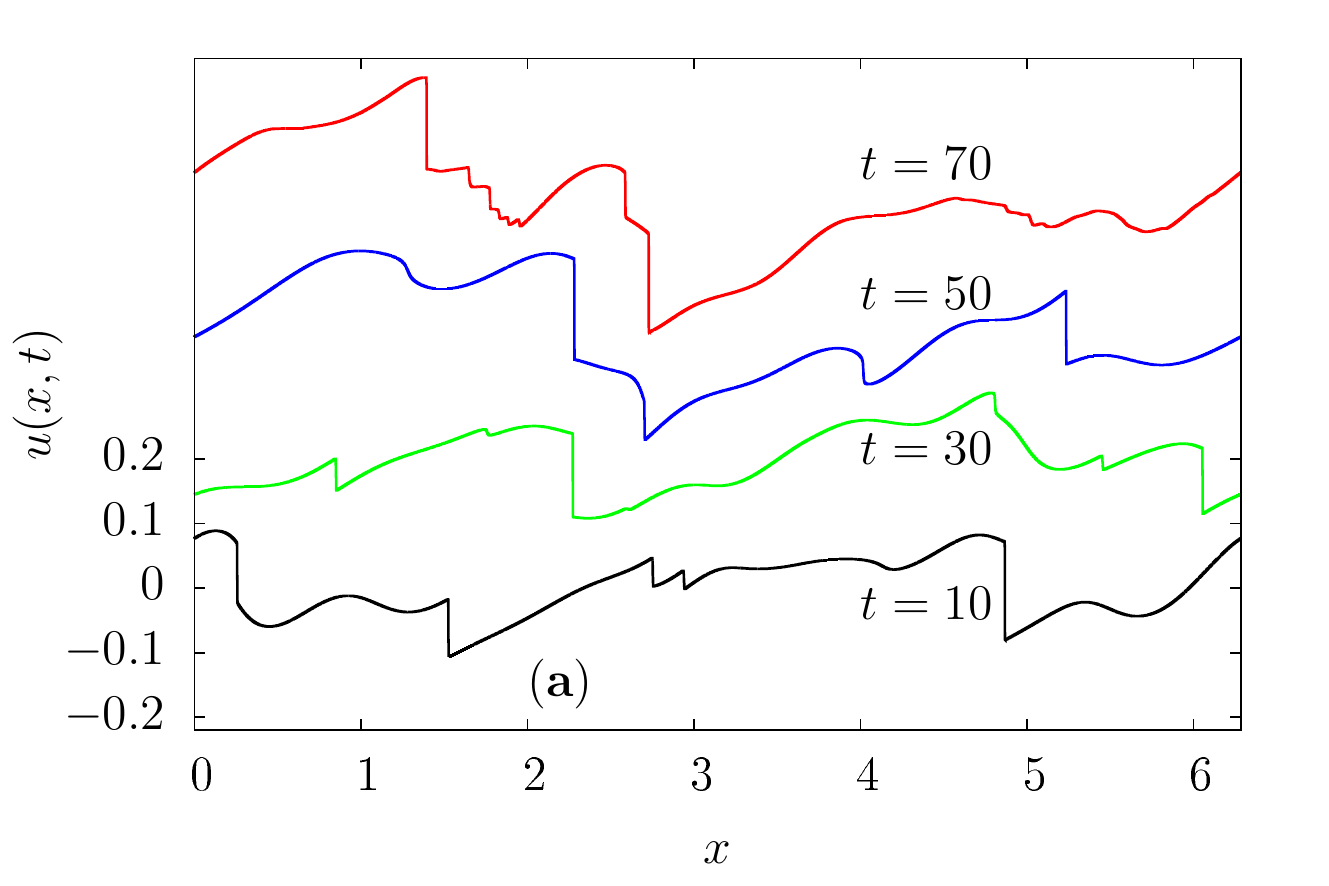}
\includegraphics[width=0.497\linewidth]{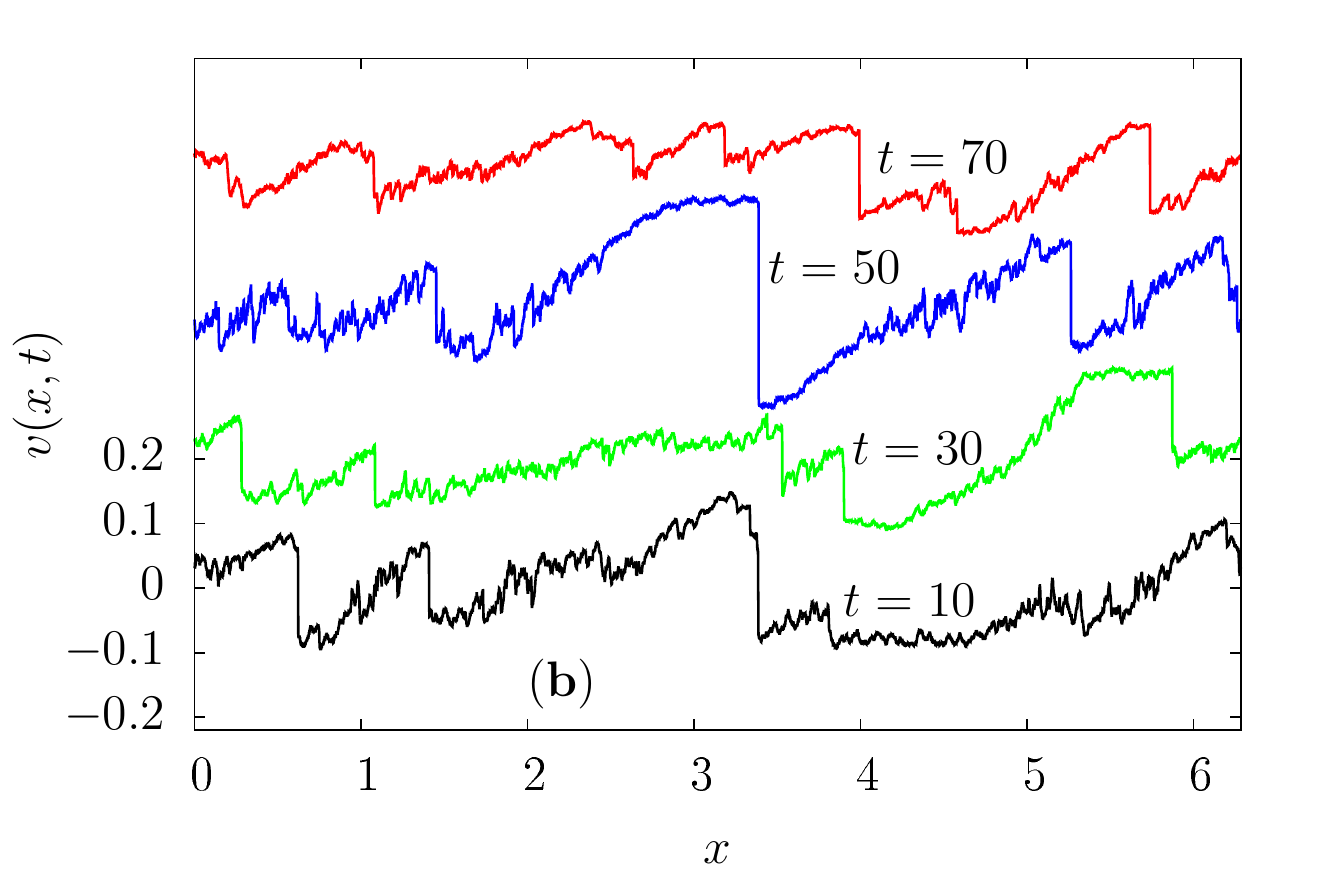}
\caption{\label{fig:time_evolve} (Color online) Representative plots of the solutions  
of the stochastically forced (a) Burgers equation and (b) the decimated Burgers 
equation for fractal dimension $D = 0.99$ at times $t = 10$, $t = 30$,  
$t = 50$, and $t = 70$ (respectively from lower black (black) line to upper red (gray) line). The velocities at different time are shifted upward on the $y$ axes.} 
\end{center}
\end{figure*}

\section{The Burgers equation on a fractally decimated Fourier set}
\label{sec:Description}

Following Ref.~\cite{frisch2012} we define the fractal Fourier decimation operator $P_D$
acting on a generic field $u(x,t) =\sum_{k} e^{i k x} \hat {u}_{k}(t)$ as:
\begin{equation} 
v(x,t) = P_D u(x,t) =\sum_{k} e^{i k x}\theta_{k} \hat {u}_{k}(t) \ ,  
\end{equation} 
where $\theta_{k}$ are independently chosen
random numbers, with $\theta_{k} = \theta_{-k}$, such that $\theta_{k} = 1$,
with probability $h_k$ and $\theta_k = 0$, with probability $1-h_k$. By choosing
$h_k = k^{D-1}$, with $0<D\le1$, we introduce a quenched disorder
that suppresses, randomly, modes on the Fourier lattice. On
average we have $N(k)\propto k^D$ surviving modes at a distance $k$ from the
origin. Considering $u$ as the velocity field given by the solution of the forced, one-dimensional Burgers equation:
\begin{equation}
\frac{\partial u}{\partial t} + \frac{1}{2}\frac{\partial u^2}{\partial x} = \nu\frac{\partial^2 u}{\partial x^2} + f, 
\label{burgers}
\end{equation}
where $\nu$ is the viscosity, $u$ is $2\pi$ space-periodic in $x$ and $f$ is a
stochastic force acting on a few shells which drives the system to a 
statistical steady state.
We can then write the decimated Burgers equation, which gives the evolution for the decimated field $v$ as: 
\begin{equation}
\frac{\partial v}{\partial t} + \frac{1}{2}P_D\frac{\partial v^2}{\partial x} = \nu\frac{\partial^2 v}{\partial x^2} + f, 
\label{d-burgers}
\end{equation}
with initial conditions $v_0 = P_D u_0$ \cite{foot1}. When $D = 1$ we recover the usual one-dimensional  equation. It is important to notice that the fractal projector in front of the non-linear term in Eq. (\ref{d-burgers}) is fundamental to ensure that the non-linear convolution does not activate the decimated degrees of freedom during the system dynamics. 
It is interesting to note that the fractal decimation, as well as any other Galerkin truncation, preserves the inviscid conservation of the first three moments of the velocity field only.

We  performed a set of  numerical simulations of Eq.~(\ref{d-burgers}) by
changing the dimension $D$ between $0.70\le D \le 1.0$ and the number of
collocation points $N$ from $2^{16}$ to $2^{19}$. We choose the forcing
to be Gaussian and white-in-time, such that 
$\langle{\hat f}(k_1,t_1){\hat f}(k_2,t_2)\rangle = f_0|k_f|^{-1}\delta(t_1 - t_2)\delta(k_1 + k_2)$, 
acting only on a shell of wavenumbers at large scales $k_f \in
[1:5-10]$. We use a pseudo-spectral method with a 
second-order Adams-Bashforth scheme to integrate in time.  
Details of all simulations can be found in Table
\ref{tab:table1}. We note that the values of $\nu$ 
decreases with the dimension $D$ (see Table \ref{tab:table1}); indeed as the decimation becomes stronger the contribution of the non-linear advection term becomes 
weaker~\cite{frisch2012}. Hence to compare results from simulations with different values of $D$, we use smaller and smaller values of $\nu$ in order to observe a similar extension of the inertial range. 
We do not know if the decimated equations are well behaved as $\nu \rightarrow 0$ and if the system develops a dissipative anomaly leading to a stationary behavior for all $D$. This is an interesting point that will be addressed in future work.

\begin{table*}
\caption{\label{tab:table1} $D$, system dimension; $D = 1$ denotes the ordinary non-decimated Burgers equation 
[Eq. (\ref{burgers})], while $D<1$ represents the decimated system as described in 
Eq.~\ref{d-burgers}. $N$, number of  collocation points. $\%(D)$, percentage of 
decimated wave numbers, where  the first value is related to the lower resolution used 
while the second value is related to the higher one. $\nu$, value of the 
kinematic viscosity.
$k_f$, the range of forced wavenumbers. $C_f$, the mean energy injection, $\langle uf \rangle$. 
$N_{mask}$, number of different random quenched masks. 
$dt$, time step used in the temporal evolution.}

\begin{ruledtabular}
\begin{tabular}{|cccccccc|}
$D$& N &$\% \left( D \right )$& $\nu$& $k_f$ & $ C_f$ &  $N_{mask}$& $dt$\\ \hline
 1   &$2^{16} - 2^{18}$ &$0$         & $8.0  \,\,\, 10^{-5}$&$[1: 5-10]$& $0.01 - 0.05$ & $0$ & $5.5 \,\,\, 10^{-5}$     \\
 0.99&$2^{16} - 2^{19}$ &$8 - 10$ & $2.5 \,\,\, 10^{-5}$&$[1: 5-10]$& $0.01 - 0.05$ & $32$& $2.3 \,\,\, 10^{-5}$ \\
 0.97&$2^{16} - 2^{18}$ &$23 - 27$& $9.0 \,\,\, 10^{-6}$&$[1: 5-10]$& $0.01 - 0.05$ & $64$& $2.0 \,\,\, 10^{-5}$\\
 0.95&$2^{16} - 2^{18}$ &$36 - 40$& $5.0 \,\,\, 10^{-6}$&$[1: 5-10]$& $0.01 - 0.05$ & $64$& $1.7 \,\,\, 10^{-5}$\\
 0.90&$2^{16} - 2^{18}$ &$59 - 64$& $2.0 \,\,\, 10^{-6}$&$[1: 5-10]$& $0.01 - 0.05$ & $64$& $1.6 \,\,\, 10^{-5}$\\
 0.80&$2^{16} - 2^{18}$ &$83 - 87$& $8.0 \,\,\, 10^{-7}$&$[1: 5-10]$& $0.01 - 0.05$ & $96$& $1.5 \,\,\, 10^{-5}$\\ 
 0.70&$2^{16} - 2^{18}$ &$93 - 95$& $6.5 \,\,\, 10^{-7}$&$[1: 5-10]$& $0.01 - 0.05$ & $96$& $1.5 \,\,\, 10^{-5}$\\ 
\end{tabular}
\end{ruledtabular}
\end{table*}

Hereafter we analyze statistical properties for either a single quenched
realization of the decimation mask or after a further averaging over 
different realizations of the
quenched disorder. We indicate with 
$\langle \bullet \rangle$ the  average over
time for a single realization of the decimation mask; while  $ \overline{
\bullet }$ is used to denote an average over different quenched masks, where each mask acts as a projector.

It is well known that the ordinary forced Burgers equation develops several discontinuities (shocks)
connected by smooth, continuous ramps as it evolves in
time [see Fig. (\ref{fig:time_evolve}a)]. As soon as we introduce the fractal decimation projector, several sharp oscillatory structures appear in the solution for $v(x,t)$, even for mild decimation 
($D \lesssim 1$), as
seen in Fig \ref{fig:time_evolve}(b).
Such structures, although reminiscent of features ({\it tygers}) of the
Galerkin-truncated Burgers  equation \cite{ray11}, are  crucially different
because they are spatially much more delocalized.

\begin{figure*}
\begin{center}
\includegraphics[width=0.497\linewidth]{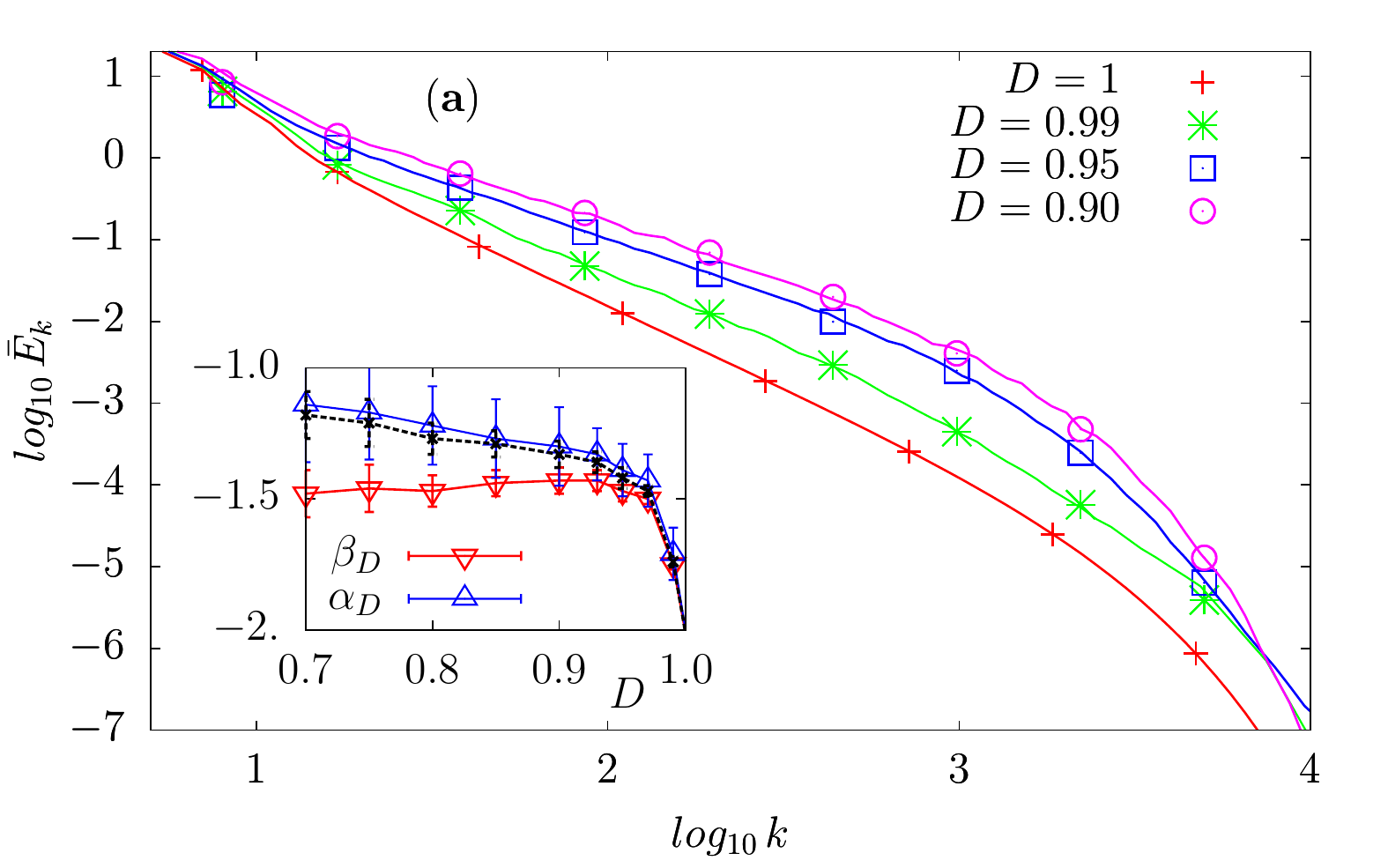}
\includegraphics[width=0.497\linewidth]{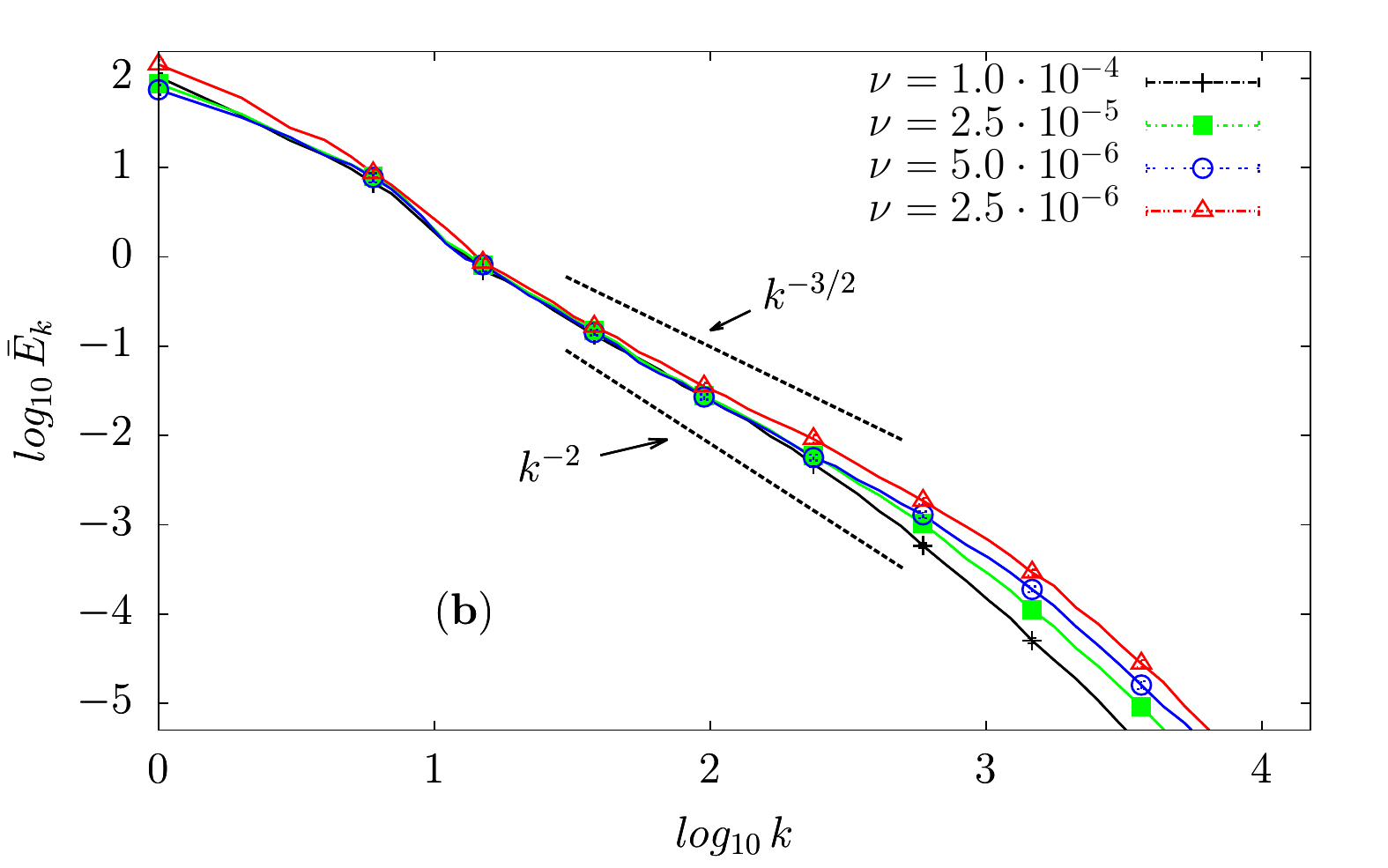}

\caption{\label{fig:spectra_expo} (Color online) (a) Mean energy spectrum,
$\bar E_k$ as a function of the wave numbers $k$; different lines and symbols
represent different fractal dimensions $D$ (see the legend for details). 
Inset: the scaling exponents $\beta_D$ (red line, downward
triangles) for the mean energy spectrum as a function of the dimension $D$; 
exponents $\alpha_D$ (blue line, upward triangles) obtained from the spectrum for 
a single projector. The latter is computed by averaging over the exponents obtained 
from each individual projector. The black dashed line confirms the relation 
$\beta_D + 1-D = \alpha_D$ as obtained in  Eq. (\ref{spect_exp}). The error bars 
for $\beta_D$ are obtained by halving the set of projectors used in the computation 
of the mean energy spectrum, while
the error bars for $\alpha_D$ are the standard deviation among all the
values used in the calculation of the mean exponent. 
(b) Mean energy spectrum (error bars are inside the symbols) at $D = 0.99$ for the following resolutions and values of the 
viscosity (see legend as well): $\nu = 1.0 \times 10^{-4}, \,\, 
N = 2^{17}$ (black line);
$\nu = 2.5 \times 10^{-5}, \,\, N = 2^{18}$ (green squares and line); 
 $\nu = 5.0 \times 10^{-6}, \,\, N = 2^{18}$ (blue circles with line); and 
$\nu = 2.5 \times 10^{-6}, \,\, N = 2^{19}$ (red triangles with line). The dashed black lines represent the scaling $k^{-3/2}$ and $k^{-2}$ as a guide to the eye.}
\end{center}
\end{figure*}

\section{Second-order Functions}
\label{sec:SecondOrder}
\subsection{Scaling in Fourier Space: The Energy Spectra}
\label{subsec:ScalingFourier}

The first question we want to address is the effect of decimation on the mean
spectral properties.  We define the energy spectrum for a general fractal
dimension as $E_k = \theta_k \langle \hat {u}^2_{k}\rangle$. Since  fractal
decimation does not break  scaling invariance of the
original equation in the inviscid limit, we still expect to observe a power-law dependency as a
function of the wavenumber $E_k \sim k^{\alpha_D}$. Another important quantity is
the spectrum averaged over the quenched disorder:  $$\bar  E_k = \overline{\langle
\hat {u}^2_{k}\rangle}  = \frac{1}{ N_{\rm mask}} \sum_{n=1}^{N_{\rm mask}}
\theta_{k}^{(n)} \, \langle \hat {u}^2_{k}\rangle,$$ where with
$n=1,\dots,N_{\rm mask}$ we indicate different realizations of the decimation
mask. From these definitions, one can infer the following relations: 
\begin{equation} 
\bar E_k \simeq k^{D-1} E_k \sim k^{\beta_D}; \,\,\, \beta_D = \alpha_D + D - 1,
\label{spect_exp} 
\end{equation} 
where we have assumed that the scaling properties of the velocity field depend only on the fractal dimension $D$ but are independent of the particular $n$-th realization of the decimation mask. To ensure the validity of Eq. (\ref{spect_exp}) we need to use $N_{mask}$ large enough to smooth out the gaps produced by the different masks in the energy spectra.

For the one-dimensional Burgers equation, because of the presence of the shock(s),
the energy spectrum scales as $\bar E_k = E_k \sim k^{-2}$ . 

In Fig. \ref{fig:spectra_expo}(a) a log-log plot of $\bar E_k$ versus $k$ for
$0.9\le D \le 1.0$ is shown.  The mean energy spectrum 
$\bar E_k \sim
k^{\beta_D}$ becomes shallower when decreasing $D$. In the inset of
Fig. \ref{fig:spectra_expo}(a), we show the dependency of the scaling exponents 
on $D$. It  changes from  $\beta_D = -2$ for
$D = 1$ to $ \beta_D = -3/2$ for $D\lesssim 0.97$, with a sharp transition
around  $ D \sim 0.97$. In the same inset we also show the validity of
the scaling relation Eq.(\ref{spect_exp}).

The existence of this quasi-singular behavior for the spectral slope at $D \sim 1$
might indicate the presence of an intermediate asymptotics spoiling the true
behavior in the limit of vanishing viscosity. To confirm this possibility, 
we show, in Fig. \ref{fig:spectra_expo}(b), the energy spectrum  $\bar E_k$ versus $k$ for $D=0.99$ at various values of the viscosity and resolution. It is seen that there is a trend, when decreasing viscosity, towards the development of an inertial range scaling $\bar E_k ~\sim k^{-3/2}$.
This suggests that, asymptotically, as $\nu \to 0$, the true scaling exponent $\beta_D
\to -3/2$ for any $D < 1$.  This observation would imply that the continuous
transition seen in the insets of Fig. \ref{fig:spectra_expo}(a) is a consequence
of some intermediate asymptotics and that 
Fourier decimation is a  singular perturbation for the spectral scaling
properties. 
\begin{figure*}
\begin{center}
\includegraphics[width=0.497\linewidth]{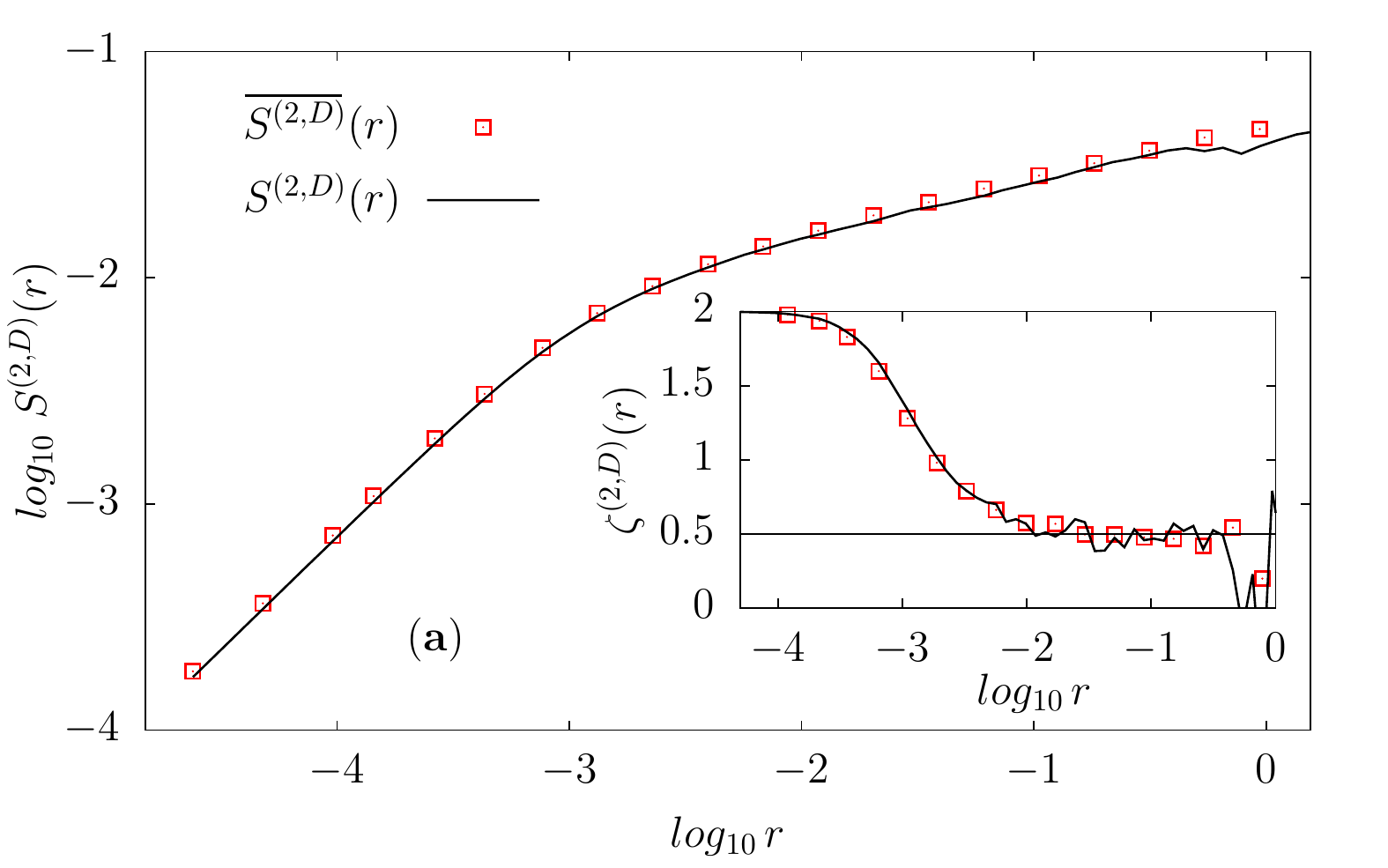}
\includegraphics[width=0.497\linewidth]{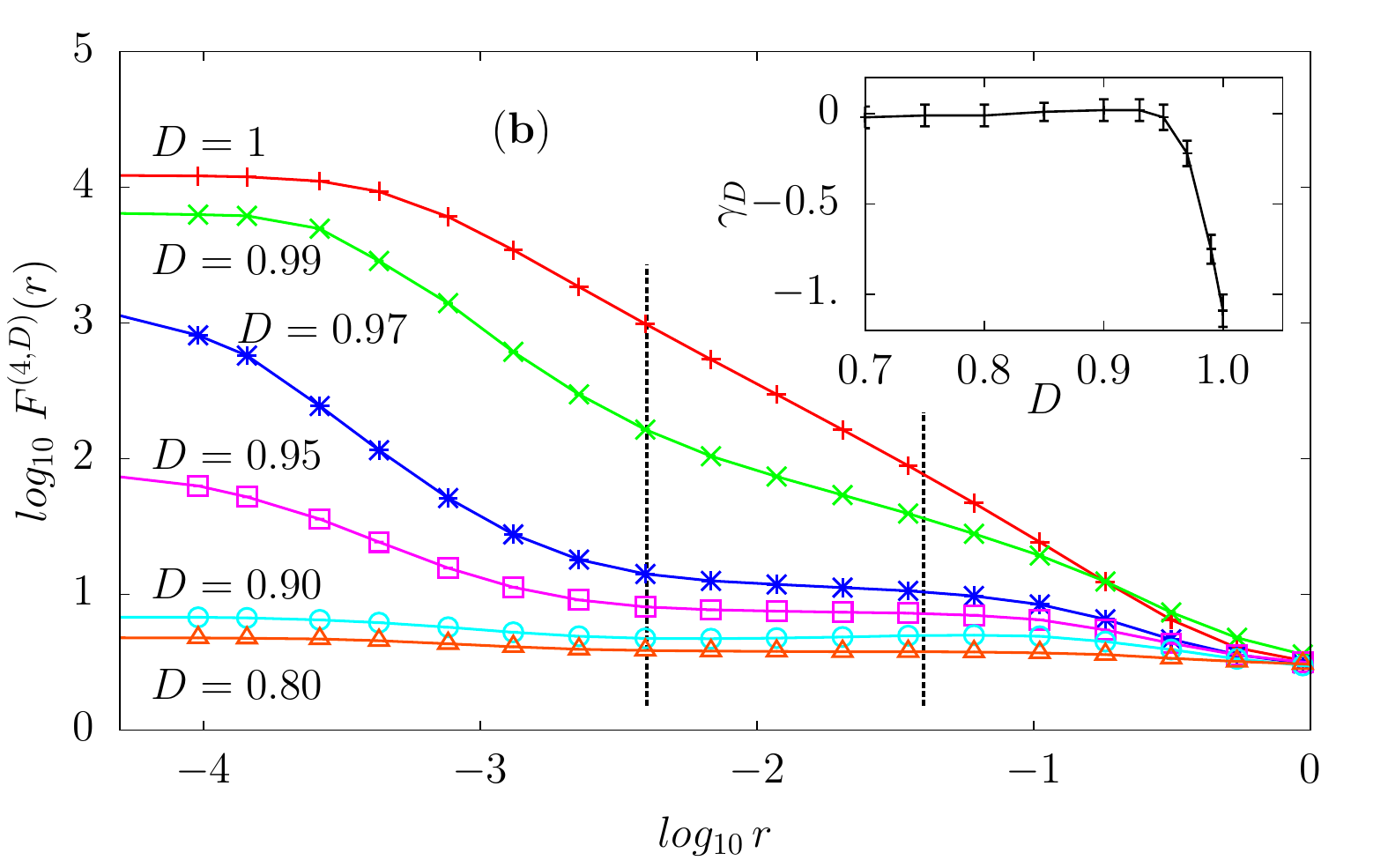}
\caption{\label{fig:Flat_and_S2} (Color online) (a) $\overline{S^{(2,D)}}(r)$ (red squares) and 
$S^{(2,D)}(r)$ (black solid line) measured at $D = 0.80$ and (inset) their 
associated local slopes [Eq. (\ref{eq:loc_slop})].  
(b) The flatness, $F^{(4,D)}(r)$ versus $r$ (log-log scale) for different dimensions $D$.
The fit is done in the range: $0.004 \le r \le 0.04$ as illustrated by the two 
vertical dotted lines to yield the exponent $\gamma_D$ as shown in the inset.}
\end{center}
\end{figure*}

\subsection{Scaling in physical space} 
\label{subsec:ScalingReal}
In order to substantiate the relation between the change in the spectrum and
the suppression of shocks  we  analyze the physical space velocity field. This is
also required to address the issue of anomalous scaling  due to the lack of a
clear definition of intermittency in Fourier space. 

\begin{figure*}
\begin{center}
\includegraphics[width=0.497\linewidth]{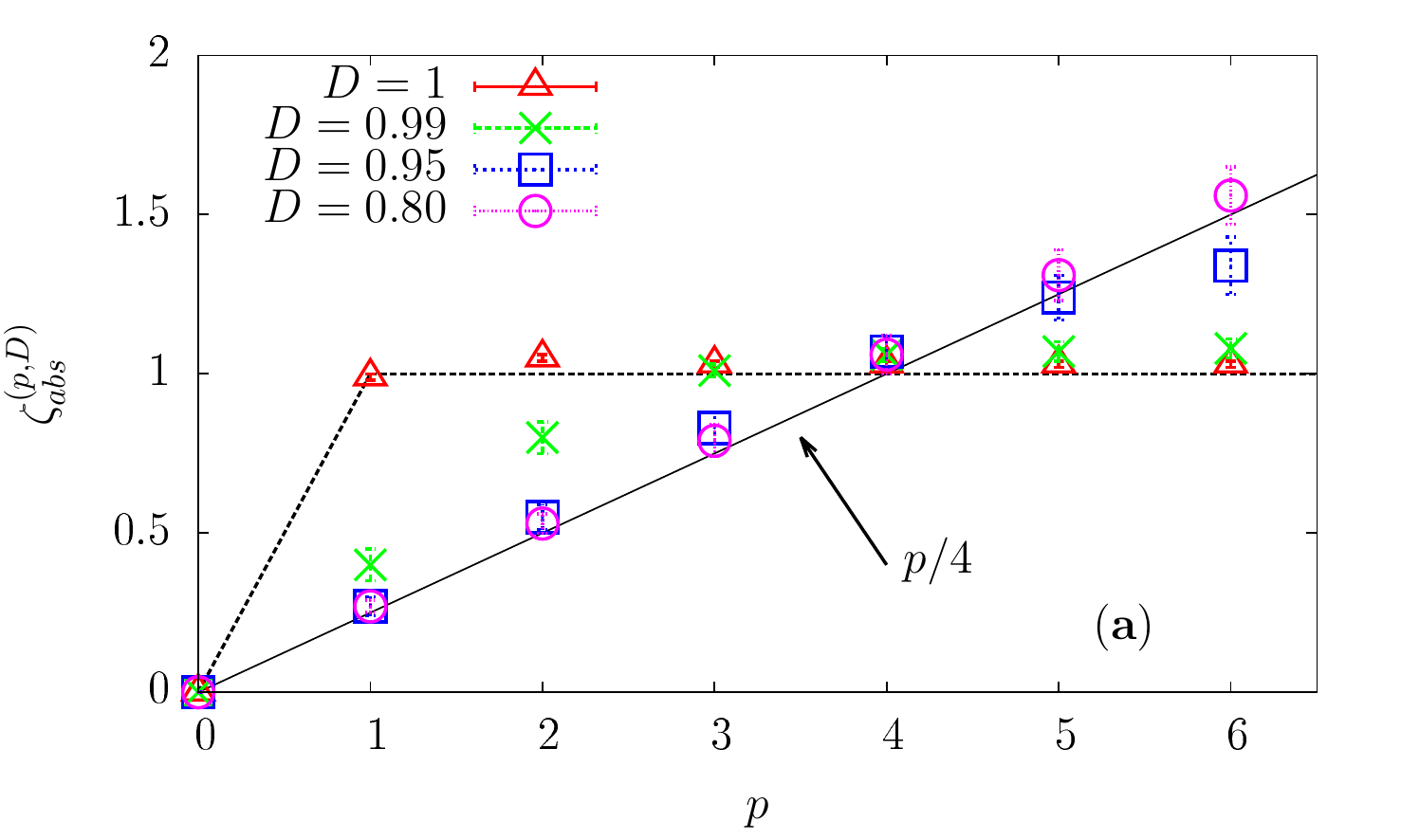}
\includegraphics[width=0.497\linewidth]{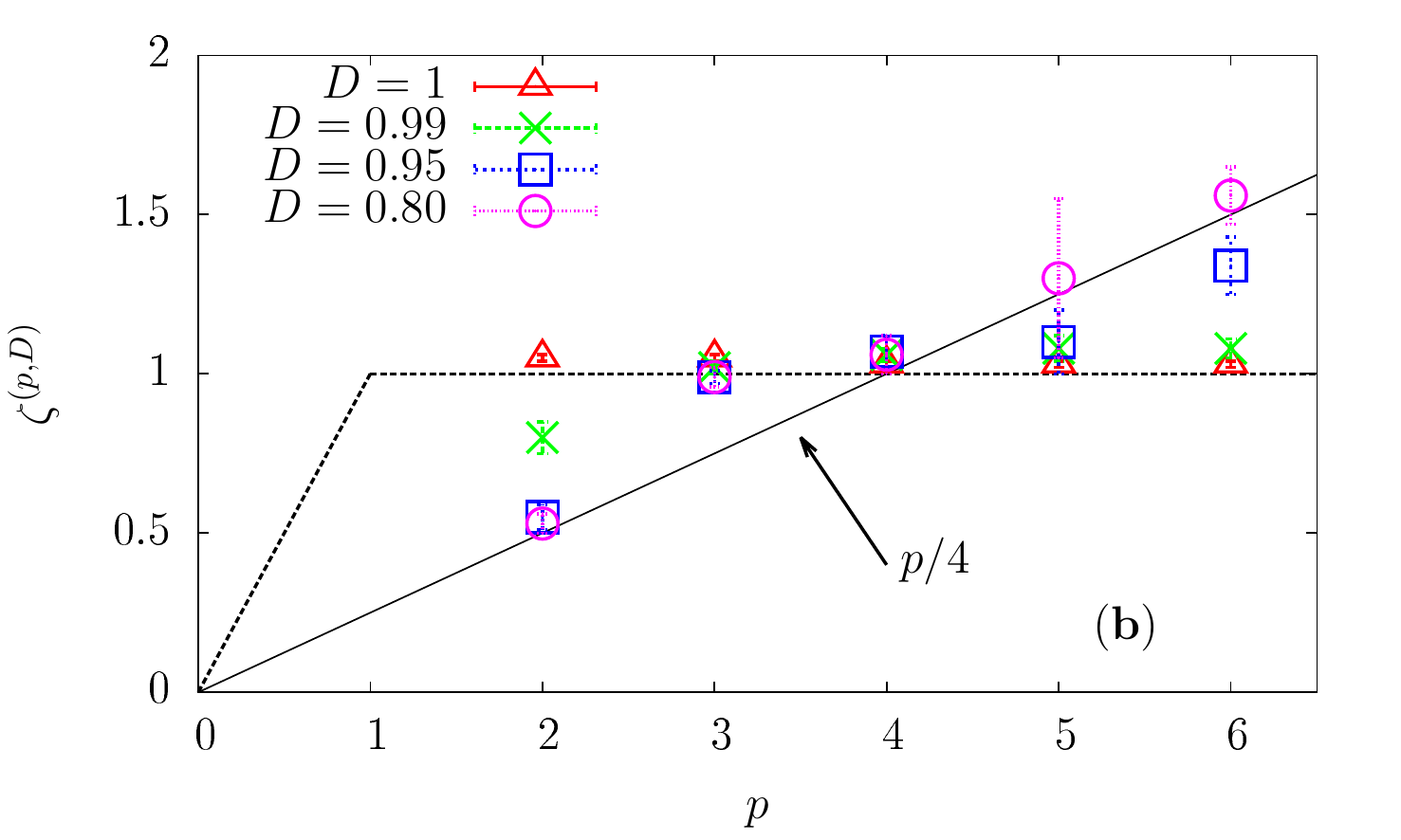}
\caption{\label{fig:Sf_exp} (Color online) 
Inertial range scaling exponents (a) $\zeta_{abs}^{(p,D)}$ and (b) $\zeta^{(p,D)}$
for the structure functions (a) with and (b) without the use of absolute values 
versus the order $p$; the different symbols are related to different
dimensions $D$ (see legend), the dashed lines are the bi-fractal behavior of the $1D$ Burgers equation. The values of the exponents are  
estimated as the best fit of local scaling exponents and the error bars are 
estimated from the variations of the local scaling exponents within the fitting range. 
We note that $\zeta_{abs}^{(3,D)}$ does not satisfy the $1D$ K\'arm\'an-Howarth analytical 
relation because of the competition between the leading and sub-leading terms 
introduced by the decimation in the scaling of the velocity field; in contrast the 
Karman-Howarth analytical relation is satisfied for the case without the absolute 
values (see text).}
\end{center}
\end{figure*}
Intermittent features in turbulent flows are quantified by the statistics of
multi-points correlation function or the so called structure functions: $$S^{(p,D)}(r)
= \langle\langle  \delta_r v  ^p \rangle\rangle_{x,t},$$ where $\delta_r v = v(x+r)-v(x)$ and  
the angular brackets, $\langle\langle \bullet \rangle\rangle_{x,t}$, denote space and time averaging over the statistically
stationary state. It is important to remark that the spatial average is 
equivalent to an average over the quenched disorder. 
To prove this we notice that:
\begin{widetext}
\begin{equation}
\overline{S^{(2,D)}}(r) = \int dx \, (\bar{v}(x+r)-\bar{v}(x))^{2} =\int dk \, (e^{ikr}-1) \bar{E_k}  =  \frac{1}{N_{mask}} \sum_{n=1}^{N_{mask}}
\int dk \, (e^{ikr}-1) \theta_k^{(n)} E_k =  S^{(2,D)}(r), 
\label{mn_2_strfct}
\end{equation}
\end{widetext}
where we have used the scaling properties of the probability defining the decimation mask and 
the fact that $dk \, \theta^{(n)}_k \stackrel{\rm \footnotesize law}{\sim} dk \, k^{D-1}$, where the symbol $\stackrel{\rm \footnotesize law}{\sim}$ stands for statistically ``in law'', i.e., the two sides have the same scaling properties when averaged on different realizations of the fractal projector.
This  relation is validated in Fig. \ref{fig:Flat_and_S2}(a)
where the second-order structure functions obtained both from a single mask (continuous 
black line) and from an average over different realizations of the quenched 
masks (square symbols in red) are shown to be identical. 
For this reason henceforth, we stop distinguishing between $\bar S_{p}(r)$ and $S_{p}(r)$.
To understand the effects of decimation on intermittency, we measure the
flatness of structure functions:

\begin{equation}
F^{(4,D)}(r) = \frac{ S^{(4,D)}(r)}{[S^{(2,D)}(r)]^2} \sim r^{\gamma_D}
\end{equation}
as a function of $r$ for different values of $D$. 
Let us stress that for the  1D Burgers equations, phenomenological and theoretical arguments \cite{becreview} predict $\zeta(p) = min(p,1)$ (see Fig. 2), which give for the flatness the scaling $r^{-1}$, [Fig. \ref{fig:Flat_and_S2}(b), inset]. As shown in Fig. \ref{fig:Flat_and_S2}(b) we find that scaling exponents, $\gamma_D$,
present the same sharp transition already observed in the slope of the energy spectra
 for $1\le D \le 0.97$ [Fig. \ref{fig:spectra_expo}(a), inset]. Thus, 
surgeries on the Fourier space and dimensional reduction seem to suppress 
intermittency in hydrodynamics (as has  also
been seen in Ref. \cite{luca2015}). We cannot refrain from  noticing that this seems to be in contrast 
with the usual phenomenology of cascade dynamics, built in terms of local-Fourier
interactions. 

\section{Higher Order Statistics}
\label{sec:StrucFunc}

\noindent The results obtained in the previous section lead us to address 
the question of whether intermittency is
indeed washed out by any small perturbation of the Fourier dynamics -- bad news
for modeling -- or if it is masked by new leading fluctuations introduced by the
modified non-linear dynamics. To answer this question, we
perform a systematic analysis of the scaling properties of structure
functions by changing the fractal Fourier dimension $D$. It is important to decompose the structure function into contributions from
the negative and positive increments of the velocity field \cite{becreview}.
We thus define $$S^{(p,D)}_+(r) = \langle\langle (\delta_r^+v)^p \rangle\rangle_{x,t}\, ; \,\, S^{(p,D)}_-(r) = \langle\langle(\delta_r^-v)^p\rangle\rangle_{x,t},$$ where $\delta_r^+v \equiv
\delta_r v > 0$ and $\delta_r^-v  \equiv \delta_r v < 0$, whence the structure
function $S^{(p,D)}(r) = S^{(p,D)}_+(r) + (-1)^p S^{(p,D)}_-(r) $.  To improve the statistics, odd-order structure functions are often
measured in terms of the  absolute value of  velocity increments; in this
case we will obviously have $S^{(p,D)}_{abs}(r) = S^{(p,D)}_+(r)  +
S^{(p,D)}_-(r) $.  To study the scaling properties it is customary to
analyze logarithmic local slopes:

\begin{eqnarray}
 \label{eq:loc_slop_abs}
\zeta^{(p,D)}_{abs}(r) &=& \frac{d \log S_{abs}^{(p,D)}(r)}{d \log r };\\
\zeta^{(p,D)}(r) &=& \frac{d \log S^{(p,D)}(r)}{d \log r }.
 \label{eq:loc_slop}
\end{eqnarray} 
The scaling exponents of order $p$ in the
inertial range are obtained as a best fit to the local exponents in the 
interval of scales where they are close to a constant.  In Fig.~\ref{fig:Sf_exp} we show
the result for both  $\zeta_{abs}^{(p,D)}$ and $\zeta^{(p,D)}$ (see figure
captions for details).

\begin{figure*}
\begin{center}
\includegraphics[width=0.49\linewidth]{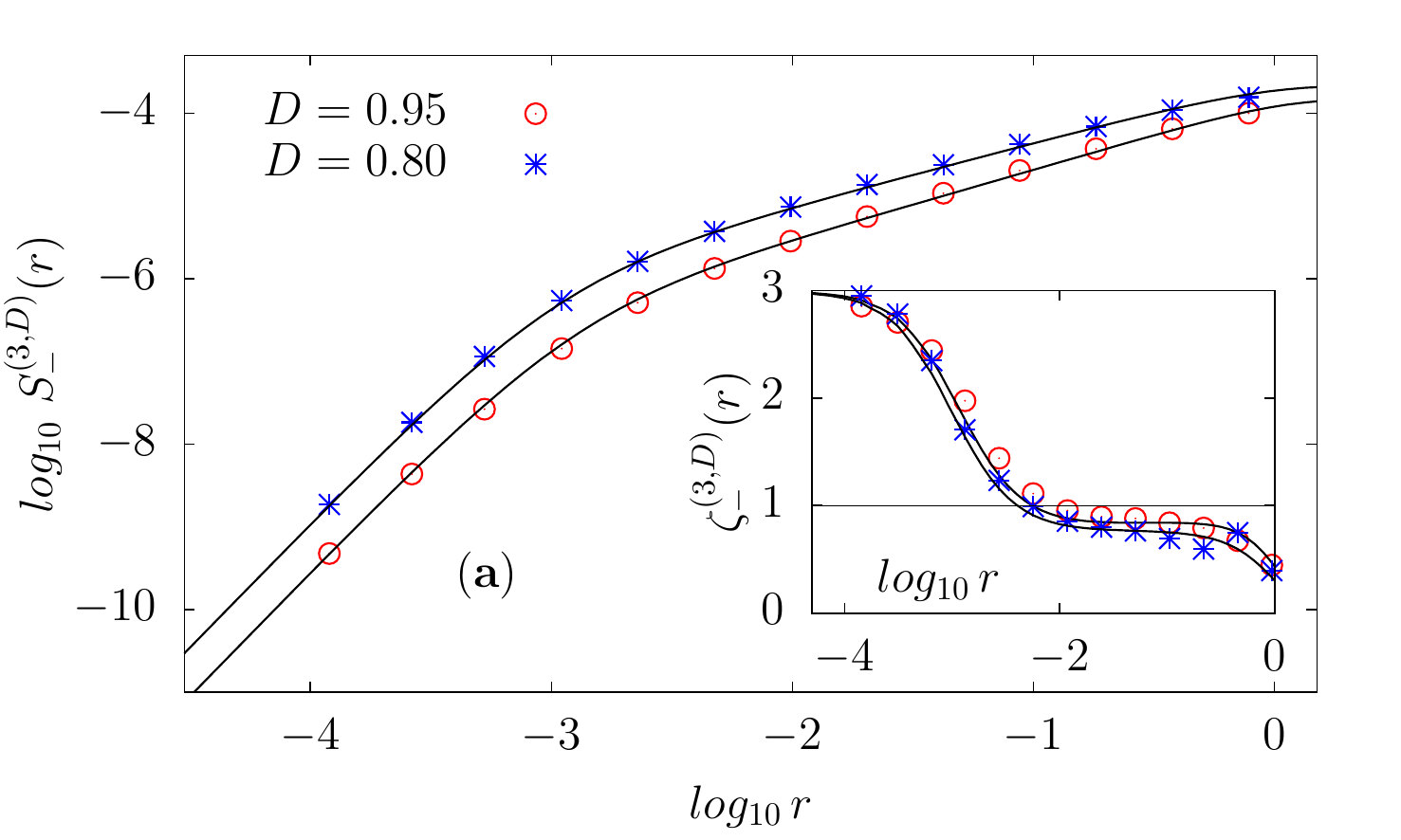}
\includegraphics[width=0.49\linewidth]{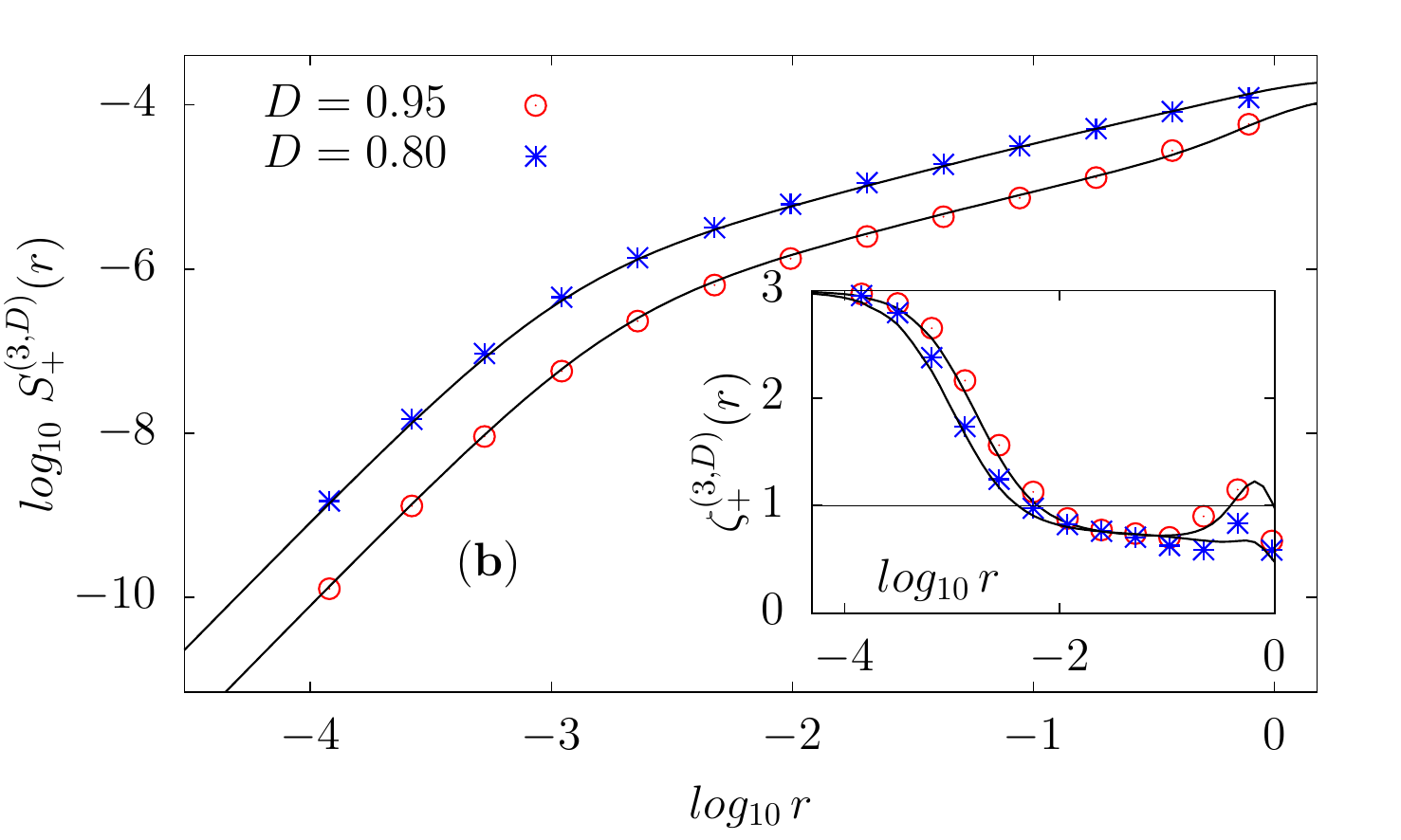}
\caption{\label{fig:fitting_Sf} (Color online) (a) Structure functions for 
$D = 0.80$ (blue stars) and $D= 0.95$ (red circles) for (a) the negative 
increments $S_-^{(3,D)}(r)$ and (b) positive increments structure functions 
$S_+^{(3,D)}(r)$; the black solid lines are the respective fitting functions [Eq.
(\ref{eq:interpola})]. In the insets, with the same legend, we show the associated 
local slopes.}
\end{center}
\end{figure*}
From a comparison of the two figures one can conclude a few important facts.
First, there is a clear tendency for even-order moments to approach the
non-intermittent scaling behavior with exponent $p/4$ as soon as $D <1$;  the
agreement being almost perfect already at $D \le 0.95$. Second, the odd-order
moments of the structure functions, defined without absolute values, seem to
maintain a memory of the original non-decimated Burgers behavior, namely $\zeta(p,D) = 1\, \forall p
\ge 1$, even for small fractal dimensions $D \ll 1$. Let us note that
this behavior is not entirely unexpected. First, the emergence of a linear scaling $p/4$ is in agreement with the observation of the spectral slope  $\beta_D
= -3/2$ and with the absence of intermittency. Figure \ref{fig:Sf_exp}(a) is
thus a demonstration of the suppression of intermittency in Burgers flows under
fractal Fourier decimation, similar to what  has been observed for the
Navier-Stokes evolution in Ref. \cite{luca2015}. The deviation from the values $p/4$ for the odd-order moments is explained by noticing that the third-order structure
function must satisfy an analytical relation similar to the K\'arm\'an-Howarth
$4/5$ law of Navier-Stokes, namely:
$S^{(3,D)}(r) = - 6 \epsilon r$ for all $D$ and where $\epsilon$ is the mean
energy dissipation. Indeed, Fig. \ref{fig:Sf_exp}(b) clearly
supports this statement. A possible way to rationalize these apparently
contradictory results is to suppose that decimation introduces a distributed noise at all scales, leading to  a typical scaling $\delta_r v \propto r^{1/4}$ on top of an underlying Burgers-like 
dynamics. If this is true, it should be detectable by looking separately at positive or
negative velocity increments. As a result, we suggest the presence of two different asymptotics:
\begin{equation}
\begin{cases}
S^{(p,D)}_+(r) =  r^{p/4} + {\rm smooth};\\
S^{(p,D)}_-(r) =  r^{p/4} +  r + {\rm smooth}, 
\label{eq:twofit}
\end{cases}
\end{equation}
where the first term on the right hand side of the equations  
should have pre-factors that go to zero for $D \to 1$. 
In Eq. (\ref{eq:twofit}) the Burgers scaling $\propto r$ is present only for the negative 
increments and {\it smooth} denotes the 
sub-leading differentiable terms induced by the viscous contribution $\propto r^p$. 

\begin{figure*}
\begin{center}
\includegraphics[width=0.49\linewidth]{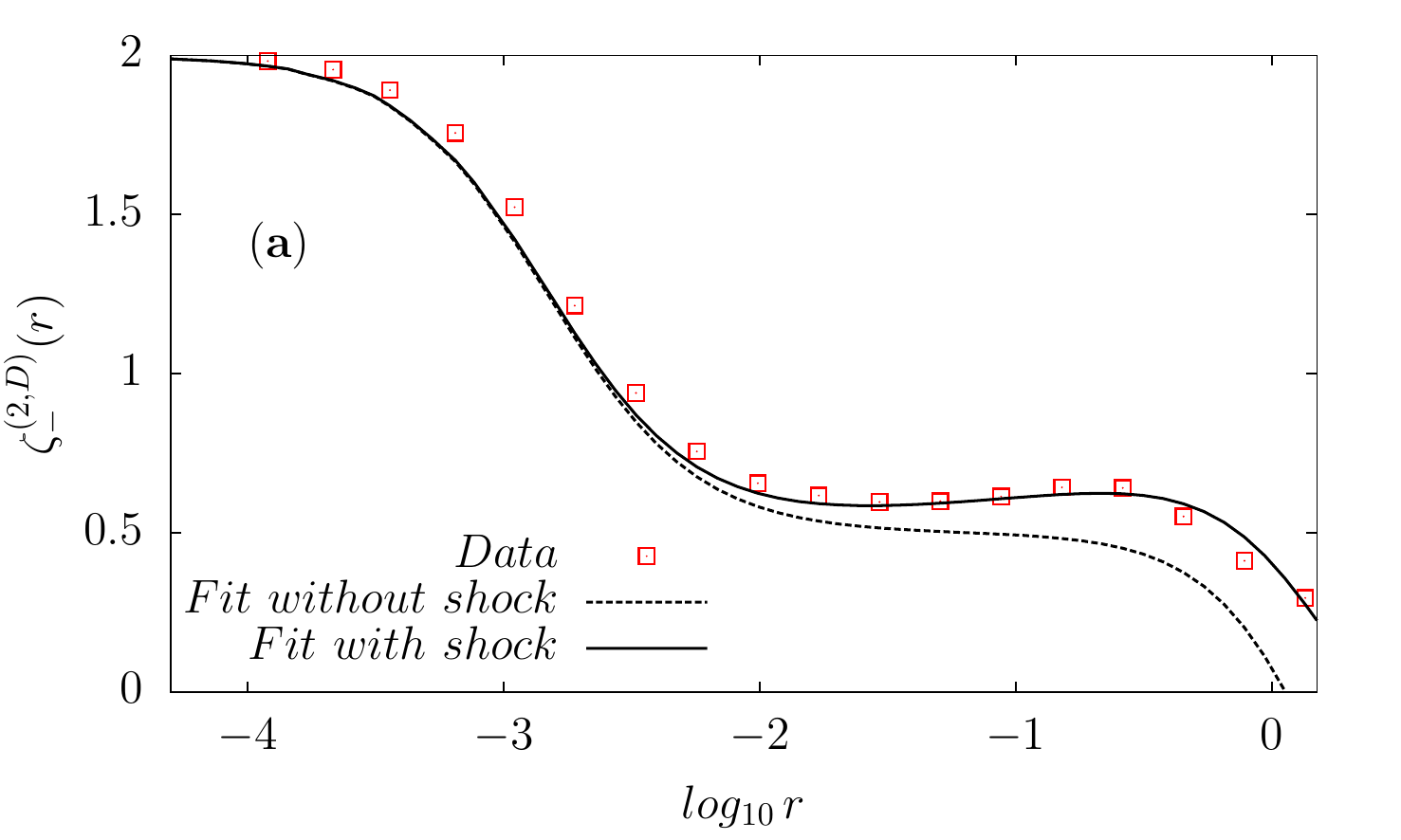}
\includegraphics[width=0.49\linewidth]{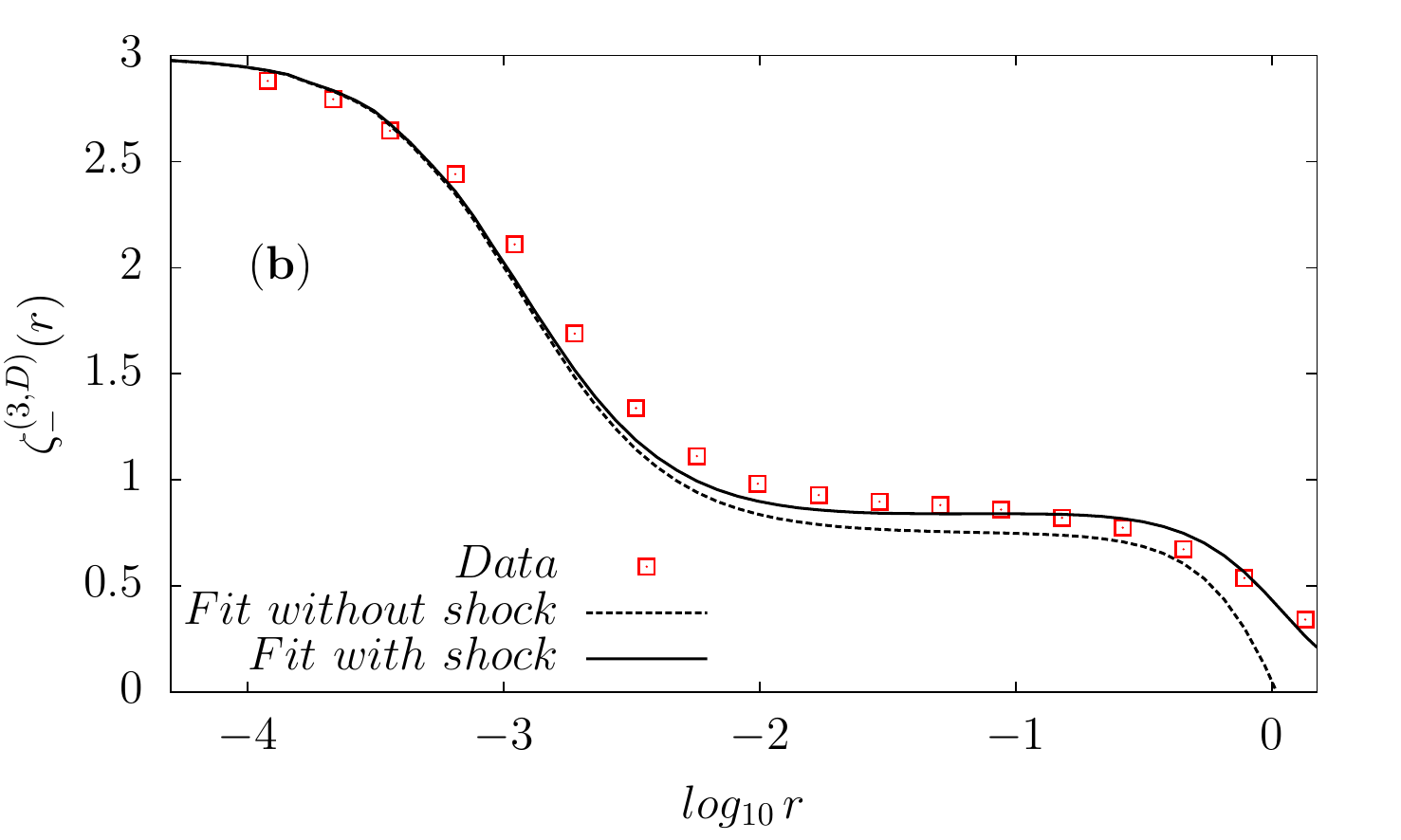}
\caption{\label{fig:fit_no_shock} (Color online) (a)
Local slopes of the structure functions associated with the negative increments at 
$D = 0.95$ for (a) the second order [$S_-^{(2,D)}(r)$] and (b) the third 
order [$S_-^{(3,D)}(r)$]; the black solid lines are obtained from the fitting function
[Eq. (\ref{eq:interpola})] and the dashed lines from the fit obtained without the shock 
contribution by setting $B^{(p,D)}_{+,-} = 0$ in [Eq. (\ref{eq:interpola})]. }
\end{center}
\end{figure*}

\begin{figure}
\begin{center}
\hspace*{-0.6cm} 
\includegraphics[width=1.1\linewidth]{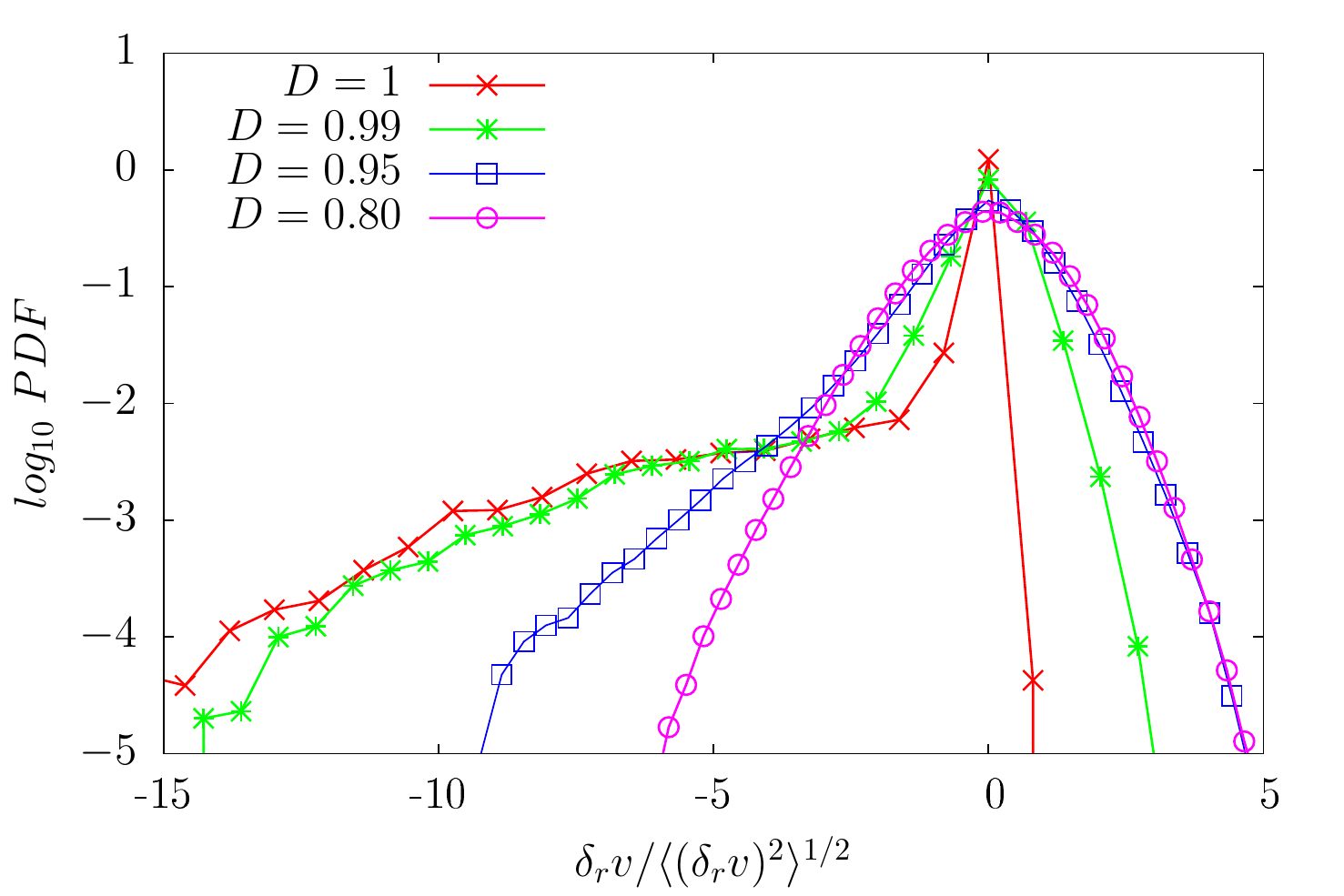}
\caption{\label{fig:pdf} (Color online) The probability density function (PDF) of 
the velocity increments at a scale $r \sim 0.005$; the PDF is normalized by 
its standard deviation. The different lines correspond to different dimensions $D$ as 
shown in the legend.} 
\end{center}
\end{figure}

In order to quantitatively check the above prediction, we perform a
series of systematic fits to $S^{(p,D)}_{\pm}(r)$ by using the following
interpolation expression for the asymptotic behavior (\ref{eq:twofit}):
\begin{widetext}
\begin{equation} 
\begin{cases} 
S^{(p,D)}_+(r) &= \left ( \frac{ A_+^{(p,D)}
r^p}{(r^2 + \eta^2)^{(p-p/4)/2}} + B_+^{(p,D)}r^p \right ) \left ( 1 +
\frac{r}{L}\right)^{c_+};\\ 
S^{(p,D)}_-(r) &=  \left (\frac{ A_-^{(p,D)}
r^p}{(r^2 + \eta^2)^{(p-p/4)/2}}  + \frac{ B_-^{(p,D)} r^p}{(r^2 +
\eta^2)^{(p-1)/2}} \right ) \left ( 1 + \frac{r}{L}\right)^{c_-},
\label{eq:interpola} 
\end{cases} 
\end{equation} 
\end{widetext}
where $A^{(p,D)}_{\pm},B^{(p,D)}_{\pm}$ are fitting constants and $\eta$ is the
dissipative scale such that for $r < \eta$, $S^{(p,D)}(r) \propto  r^p$.
The overall factor $(1 + r/L)^{c_{\pm}}$ is used to saturate the inertial 
range scaling for $r$ beyond the forcing length scale. $L$ and $c_{\pm}$ are estimated as the best-fitting parameters. 
From Eq. (\ref{eq:interpola}) it is easy to recognize that the first term on the 
right hand side of both equations represents the effect introduced by 
the decimation, while the second term (present only for negative increments) 
represents the standard shock-dominated Burgers scaling (plus viscous contributions). 
Clearly, if $A_+\sim A_-$ for all $D$, we have moments of even order that are dominated
by the $p/4$ scaling while moments of odd order have the usual Burgers scaling.  
In Figs. \ref{fig:fitting_Sf}(a) and \ref{fig:fitting_Sf}(b) we show the best fit by using these
expressions for the third-order moment ($p=3$) and
$D=0.95,0.8$ for both the (a) negative and (b) positive increments. 
For the case of negative increments, we also show the best fit with and without 
the shock contribution to highlight the importance of the shock to 
reconstruct the right behavior for $p=2$ in Fig.~\ref{fig:fit_no_shock}(a) and 
for $p=3$ in Fig.~\ref{fig:fit_no_shock}(b). 
As one can see, the Eqs. (\ref{eq:interpola}) are able to 
reconstruct the local scaling properties in a robust way, 
showing that our phenomenological model is {\it not incompatible} with the data. 
Finally we measure the probability density
function of the velocity increments at different scales (Fig.
\ref{fig:pdf}). This clearly shows the emergence of non-trivial fluctuations for 
positive velocity increments at decreasing fractal dimension $D$; such fluctuations  
are almost absent in the standard one-dimensional Burgers case.
In summary, all our results indicate that while on the one hand
the numerical evidence point toward a robustness of the shock 
structure (as also visually confirmed from
Fig. \ref{fig:time_evolve}), on the other hand, decimation introduces important fluctuations that spoil the scaling of the original undecimated equation without modifying  the existence of a  constant-flux of energy from large to small-scales.

\section{Conclusions}
\label{sec:Conclusions}
Let us  now turn to a few theoretical considerations
for understanding the behavior of $\zeta^{(p,D)}$.  We first recall that just like
any  Galerkin truncation~\cite{ray11}, the fractal Fourier decimation
constrains the number of conserved quantities to the first three moments of
the velocity in the Burgers equation. In particular, this allows the
conservation of a cubic moment whose relative flux would  yield $\zeta_4 = 1$,
which, in turn, would be consistent with our numerical result $\zeta_p = p/4$.
Another possible explanation for the $E(k) \propto k^{-3/2}$ scaling is  
the idea that a new decorrelation mechanism in the shell-to-shell
energy transfer across Fourier modes might be introduced by the fractal
decimation. The fractal mask can be seen as an extra, ad-hoc removal of non-linear
couplings at all scales and, as such, a sort of power-law external ``energy-conserving'' noise.  It is not unphysical to suppose that due to the power-law
dependence on the wavenumber, a different weight between
local and non-local interactions in Fourier space is introduced, making the latter more
important then in the usual $D=1$ case. Given all this, it is conceivable that an
extra decorrelation time of the order of $\tau_\mathrm{dec}(k) \propto 1/k$ appears,
leading to a slow down of the energy transfer mechanisms, as is the case for 
Alfv\'{e}n waves in MHD~\cite{IK} or in the presence of a rapid distortion 
mechanism~\cite{RDT}. Typically, this leads to an estimate for
the energy flux $\epsilon = kE(k)/\tau_\mathrm{tr}(k)$, where the transfer time is
given in terms of a golden  mean between the eddy-turn-over time
$\tau_{\mathrm{eddy}}(k)$ and the decorrelation time $\tau_\mathrm{dec}(k)$, $\tau_\mathrm{tr}(k) =
\tau_\mathrm{eddy}^2/\tau_\mathrm{dec}(k)$. If this is the case, considering that
$\tau_\mathrm{eddy} \propto (k^3 E(k))^{-1/2}$, we arrive at the estimate $E(k) \propto
k^{-3/2}$.

Let us notice that other decimation protocols might be imagined. In particular, one can consider performing a selective decimation of a single class of triads (e.g., local or nonlocal), in order to probe the main mechanisms leading to the formation of small-scales shocks in the dynamical evolution. In this case decimation cannot be univocally defined in terms of each wavenumber, i.e. one wave number might belong to a local or non-local triads depending on the other two. Hence a selected triads reduction can be done only inside the non linear convolution term, accessible via a fully spectral code with strong limitation in the numerical resolution achievable; see Ref. \cite{SmthLesl}.\\
Moreover, a recent study \cite{BMBB} has shown that highly non-trivial time correlations among Fourier triads are connected to the presence of intermittency in physical space. It is not obvious \emph{a priori} that reducing Fourier interaction will lead to a time de-synchronization of the energy exchange among triads. The results of Ref. \cite{BMBB} together with the ones shown here suggest that the build up of small-scale intermittent fluctuations in physical space (shocks)  is indeed the outcome of an entangled temporal correlations amongst many (all?) Fourier modes.
Another interesting  potentially useful methodology is to apply proper orthogonal decomposition of Fourier amplitudes and phases correlations \cite{Holmes}.

In this paper, we have presented a set of theoretical and numerical results
concerning the evolution of the one-dimensional stochastically forced Burgers
equation decimated on a fractal Fourier set. Decimation leads, very quickly, 
to a suppression of the shock-dominated statistics, indicating that the
bifractal scaling properties of the original equation are very
sensitive to the details of the dynamical evolution. Similar results have also
been recently obtained for the more complicated case of the dynamics of fully
developed incompressible turbulent flows in three dimensional Navier-Stokes equations. Some
properties connected to the existence of shock-like solutions are
nevertheless robust, but sub-leading.  Our results indicate that the existence
of strong localized fluctuations in Burgers is the result of highly entangled
correlations among {\it all} Fourier modes. This might be important to develop
models for the nonlinear evolution based on suitable reduction (and
replacement) of a subset of the original degrees of freedom. 

\acknowledgments

MB and LB acknowledge funding from the European Research Council under the
European Union's Seventh Framework Programme, ERC Grant Agreement No 339032. LB and SSR thank COST ACTION MP1305 for support. SSR acknowledges the support of the Indo-French Center for Applied Mathematics (IFCAM) and AIRBUS Group Corporate Foundation Chair in Mathematics of Complex
Systems established in ICTS and the hospitality of the Department of
Physics, University of Rome ``Tor Vergata'', Rome, Italy and the Observatoire
de la C\^ote d'Azur, Nice, France.


\begin{thebibliography}{10}
\bibitem{frischbook}  U. Frisch, {\it Turbulence: The Legacy of A.N.
Kolmogorov} (Cambridge University, Cambridge, UK, 1996).

\bibitem{K41} A.N. Kolmogorov, Dokl. Akad. Nauk SSSR
{\bf 30}, 301 (1941).


\bibitem{grisharmp} G. Falkovich, K. Gawedzki and M. Vergassola,
Rev. Mod. Phys. {\bf 73}, 913 (2001).

\bibitem{ruelle} J.-D. Fournier and U Frisch, Phys. Rev. A {\bf 28}, 
1000 (1983).


\bibitem{vassilicos} {\it Intermittency in Turbulent Flows}, ed. J. C. 
Vassilicos, (Cambridge University, Cambridge, UK, 2001).


\bibitem{astro1} J.Cho, A. Lazarian, and E. T. Vishniac, ApJ {\bf 564}, 291, (2002).

\bibitem{astro2} W.-C. Muller and D. Biskamp, Phys. Rev. Lett. {\bf 84}, 475, (2000). 

\bibitem{astro3} P. Veltri. Plasma Physics and Controlled Fusion {\bf 41},A787, (1999). 

\bibitem{astro4} R. Grauer, J. Krug and C. Marliani. Phys. Lett. A {\bf 195}, 335, (1994). 

\bibitem{geo1} J. J. Riley and E. Lindborg, J. Atmos. Sci., {\bf 65}, 2416–2424, (2008).

\bibitem{geo2} G. Boffetta and R. E. Ecke, Ann. Rev. of Fluid Mech., {\bf 44} 427, (2012).


\bibitem{math1} R. Benzi, G. Paladin, G. Parisi and A. Vulpiani. Journal of Physics A: Mathematical and General, {\bf 18(12)}, 2157 (1985).

\bibitem{math2} A. C. Newell, S. Nazarenko and L. Biven, Physica D: Nonlin. Phenom. {\bf 152-153}, 520 (2001).

\bibitem{math3} Chorin, A. Joel. Commun. Pure Appl. Math. {\bf 34}, 853 (1981).

\bibitem{math4} U. Frisch. and G. Parisi, \emph{Turbulence and Predictability in Geophysical Fluid Dynamics and Climate Dynamics}, edited by M. Ghil, R. Benzi, and G. Parisi (North-Holland, Amsterdam, 1985), Vol. 88, pp 71-88.


\bibitem{burgulence} U. Frisch and J. Bec,
{\it Les Houches 2000: New Trends in Turbulence}, Eds. : M. Lesieur, A. Yaglom and F. David, pp. 341-383, Springer EDP-Sciences, (2001).

\bibitem{becreview} J. Bec and K. Khanin, Phys. Rep. {\bf 447}, 1-66, (2007).


\bibitem{frisch2012} U. Frisch, A. Pomyalov, I. Procaccia, and S. S. Ray, 
Phys. Rev. Lett. {\bf 108}, 074501, (2012).

\bibitem{rayreview} S. S. Ray, Pramana J. of Phys., {\bf 84}, 395, (2015).

\bibitem{luca2015} A. S. Lanotte, R. Benzi, S. K. Malapaka, F. Toschi and L. Biferale. Phys. Rev. Lett.  {\bf 115}  264502 (2015)

\bibitem{foot1} The forcing is also projected on the decimated lattice.

\bibitem{ray11} S. S. Ray, U. Frisch, S. Nazarenko, and T. Matsumoto, 
Phys. Rev. E {\bf 84}, 016301, (2011).

\bibitem{IK} P. S. Iroshnikov, Sov. Astron. {\bf 7}, 566 (1964); 
R. H. Kraichnan, Phys. Fluids {\bf 8}, 1385, (1965).

\bibitem{RDT} J. C. R. Hunt and D. J. Carruthers, J. of Fluid Mech. 
{\bf 212}, 497, (1990).

\bibitem{SmthLesl} Smith, M. Leslie and Y. Lee, Journal of Fluid Mechanics {\bf 535}, 111-142, (2005).

\bibitem{BMBB} M. Buzzicotti, B. P. Murray, L. Biferale, M. D. Bustamante, arXiv:1509.04450, (2015).

\bibitem{Holmes} P. Holmes, J. L. Lumley \& G. Berkooz, \emph{Turbulence, coherent structures, dynamical systems and symmetry}. Cambridge university press, (1998).

\end{thebibliography}
\end{document}